%% file: mainText.tex
\documentclass[
reprint,
superscriptaddress,
 amsmath,amssymb,
prb,
floatfix,
]{revtex4-2}

\usepackage{graphicx}% Include figure files
\usepackage{dcolumn}% Align table columns on decimal point
\usepackage{bm}% bold math
\usepackage{xcolor}
\usepackage{upgreek}

\newcommand{\affilMSNE}{\affiliation{Department of Materials Science and NanoEngineering, Rice University, Houston, TX 77005, USA}}
\newcommand{\affilECE}{\affiliation{Department of Electrical and Computer Engineering, Rice University, Houston, TX 77005, USA}}
\newcommand{\affilIMR}{\affiliation{Institute for Materials Research, Tohoku University, Sendai 980-8577, Japan}}
\newcommand{\affilCHBE}{\affiliation{Department of Chemical and Biomolecular Engineering, Rice University, Houston, TX 77005, USA}}
\newcommand{\affilPA}{\affiliation{Department of Physics and Astronomy, Rice University, Houston, TX 77005, USA}}
\newcommand{\affilSCI}{\affiliation{Smalley-Curl Institute, Rice University, Houston, TX 77005}}
\newcommand{\affilRAMI}{\affiliation{Rice Advanced Materials Institute, Rice University, Houston, TX 77005}}

\begin{document}

\preprint{APS/123-QED}

\title{Unconventional Temperature Dependence of Exciton Diamagnetism in 2D Ruddlesden-Popper Lead Halide Perovskites}% Force line breaks with \\
% \thanks{A footnote to the article title}%

\author{William A. Smith}
    \affilMSNE
\author{Fumiya Katsutani}%
    \affilECE
\author{Jin Hou} % I thought Jin was the sample grower, why was Hao ahead of him on the author list?
    \affilCHBE
\author{Hao Zhang} % Clarify his contribution
    \affilCHBE
% \author{Satoshi Matsuzawa} % Who is this?
%     \affilIMR
% \author{Hiromasa Yasumura} % Who is this?
%     \affilIMR
\author{Jean-Christophe Blancon}
    \affilCHBE
\author{Hiroyuki Nojiri}
    \affilIMR
\author{Aditya D. Mohite}
    \affilCHBE
    \affilSCI
\author{Andrey Baydin}
    \email{ab133@rice.edu}
    \affilECE
    \affilSCI
    \affilRAMI
\author{Junichiro Kono}
    \email{kono@rice.edu}
    \affilMSNE
    \affilECE
    \affilSCI
    \affilRAMI
    \affilPA
\author{Hanyu Zhu}
    \email{hanyu.zhu@rice.edu}
    \affilMSNE
    \affilECE
    \affilSCI
    \affilRAMI
    \affilPA

\date{\today}% It is always \today, today,
             %  but any date may be explicitly specified

%%%  THE ABSTRACT
\input{Sections/0_abstract}
\maketitle

%%% MAIN TEXT

\section{\label{sec:intro}Introduction}

\input{Sections/1_introduction}

\section{\label{sec:methods}Materials and Methods}

\input{Sections/2_methods}

\section{\label{sec:results}Temperature-Dependent Diamagnetic Shift}%Results and Discussion}

\input{Sections/3_results}

\input{Sections/4_discussion}
\section{\label{sec:Conclusion}Conclusion}

\input{Sections/5_conclusion}
\\

% \section*{Author Contributions}

% Temperature-dependent magnetospectroscopy with RAMBO-I at 25T was performed by FK. 40T magnetospectroscopy in RAMBO-II was performed by WAS. Magnets, banks, and controllers for RAMBO-I and -II designed and built by HN. Sample preparation by HZ and JH. 

\begin{acknowledgments}
We acknowledge funding support from the National Science Foundation through Grant No.\ DMR-2019004 and the Robert A.\ Welch Foundation through Grants Nos.\ C-1509 and C-2128.
%This work was done in part using resources of the Shared Equipment Authority at Rice University.

\end{acknowledgments}
% \clearpage
\appendix

\section{Timing and Synchronization}
\renewcommand{\thefigure}{\thesection.\arabic{figure}}
\setcounter{figure}{0}
\input{Sections/A_Experiment}

\section{Heating and Cooling}
\renewcommand{\thefigure}{\thesection.\arabic{figure}}
\setcounter{figure}{0}
\input{Sections/A_ZeroField}

\section{Detailed Results}\label{sec:Synch}
\renewcommand{\thefigure}{\thesection.\arabic{figure}}
\setcounter{figure}{0}
\input{Sections/A_DetailedResults}

\clearpage
\bibliography{bibFile}% Produces the bibliography via BibTeX.

\end{document}

%% file: Sections/0_abstract.tex
\begin{abstract}
 Layered hybrid perovskites containing larger organic cations have demonstrated superior environmental stability, but the presence of these insulating spacers also strengthens the exciton binding energy, which contributes to reduced carrier separation. The consequences of increased binding energy on device efficiency are still not fully documented, and binding energy measurements are often conducted at cryogenic temperatures where linewidths are decreased and a series of hydrogen-like bound states can be identified, but not under ambient conditions where devices are expected to operate. In contrast to the quenching observed in 3D perovskites such as methylammonium lead iodide, where exciton binding energies are thought to decrease at higher temperatures, we present evidence for a smaller excitonic radius at higher temperatures in the $n=5$ member of butylammonium-spaced methylammonium lead iodide, (BA)$_2$(MA)$_{n-1}$Pb$_n$I$_{3n+1}$. We measured the temperature-dependent diamagnetic shift coefficient in magnetic fields up to 40\,T, which is one-third as large at room temperature as those at cryogenic temperatures. In both the ideal 2D and 3D hydrogen models, this trend would indicate that the exciton binding energy more than triples at room temperature.
 % , a surprising deviation from the conventional wisdom that the binding energy is relatively insensitive to temperature. 
 
\end{abstract}

%% file: Sections/1_introduction.tex
In photovoltaic applications, exciton formation reduces the density of photocreated free carriers, and the additional burden of dissociation before charge separation increases the probability of recombination before the excited carrier can be collected. The simplest case of dissociation occurs when the thermal energy $k_\text{B}T$ exceeds the exciton binding energy, generating free carriers at short time scales (as low as picoseconds)~\cite{chenEffectCarrierThermalization2015}. Meanwhile, in many hybrid perovskites with beneficial characteristics for applications such as ambient stability, strong excitonic behavior survives at room temperature with binding energies in the hundreds of meV~\cite{sidhikDeterministicFabrication3D2022, blanconExtremelyEfficientInternal2017, blanconScalingLawExcitons2018} -- many times the thermal energy. In these materials, dissociation may be promoted by engineering the band structure at junctions and built-in electric fields at the bulk or grain boundaries~\cite{blanconExtremelyEfficientInternal2017}. Reducing the exciton population of these materials at room temperature is important for improving the overall photovoltaic conversion efficiency, but the binding energy is difficult to measure due to the broad spectral features of exciton energy levels, which are often difficult to distinguish from those of the band continuum or defects~\cite{yamadaPhotoelectronicResponsesSolutionProcessed2015,niedzwiedzkiExcitonBindingEnergy2022,hansenMeasuringExcitonBinding2024a}. Often, the binding energy is measured at cryogenic temperatures with more prominent spectroscopic signatures~\cite{miyataDirectMeasurementExciton2015, yamadaPhotoelectronicResponsesSolutionProcessed2015,kazimierczukGiantRydbergExcitons2014, blanconScalingLawExcitons2018}, with the assumption that the binding energy only has a small or inverse temperature dependence, such as for various 3D perovskites~\cite{evenAnalysisMultivalleyMultibandgap2014, sestuAbsorptionFSumRule2015, niedzwiedzkiExcitonBindingEnergy2022, miyataDirectMeasurementExciton2015, yangUnravelingExcitonBinding2017}. %In typical inorganic semiconductors, this assumption is reasonable according to a simple hydrogen model for the excitons, in which the effective mass of the carriers and the dielectric function are nearly constant. 

Here, we present a counterexample to this expectation. We measured the temperature-dependent diamagnetic shift in the Ruddlesden-Popper hybrid perovskite (BA)$_2$(MA)$_{n-1}$Pb$_n$I$_{3n+1}$ (BA: butylammonium $\mathrm{CH_3(CH_2)_3NH_3}$, MA: methylammonium $\mathrm{CH_3NH_3}$) with $n = 5$ up to 40\,T with the second-generation Rice Advanced Magnet with Broadband Optics (RAMBO-II); see Ref.~\cite{noeTabletopRepetitivePulsed2013,tayMagnetoopticalSpectroscopyRAMBO2022} for the first-generation setup (RAMBO-I). We found a strong inverse correlation between the diamagnetic shift coefficient and temperature. The excitonic diamagnetic shift mainly comes from the orbital wavefunctions and is proportional to the size of the excitons. Therefore, in the simple hydrogen atom model, regardless of dimensionality, a smaller diamagnetic shift suggests a smaller average distance between the electrons and holes, and correspondingly, a large binding energy. More specifically, our results indicate that the exciton binding energy more than triples at room temperature compared to liquid helium temperatures. Such an unusual behavior calls for further experimental and theoretical investigations of exciton physics in hybrid perovskites.
\\\\\\\\\\\\

%% file: Sections/2_methods.tex
\subsection{\label{sec:material}Ruddlesden-Popper Hybrid Perovskites}

\begin{figure}[b]
    \centering
    \includegraphics[width=1\linewidth]{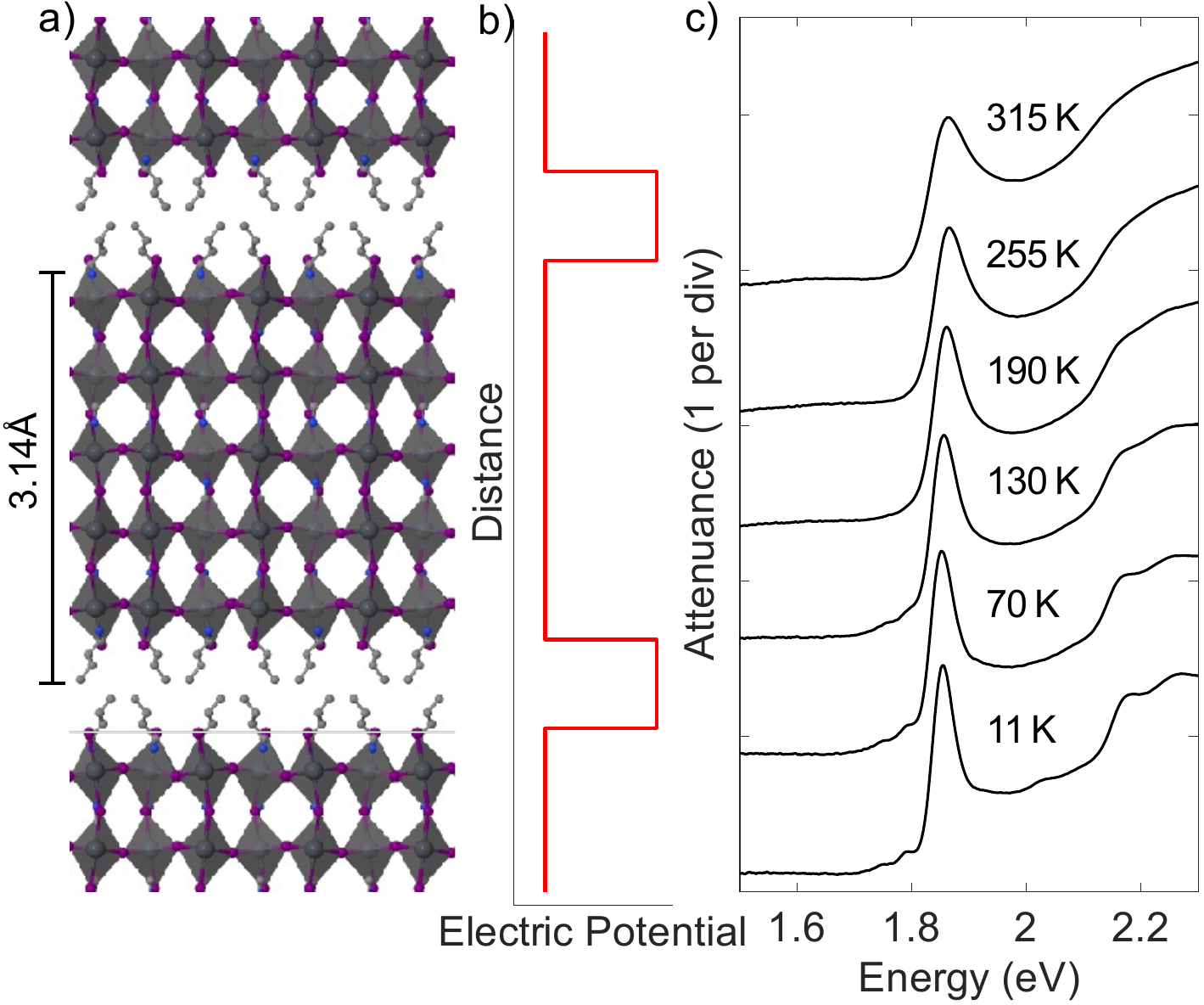}
    \caption{a)~View of $\mathrm{(BA)_2(MA)_4Pb_5I_{16}}$ crystal along [1 0 0] crystal axis. Grey: lead, purple: iodine, blue: nitrogen, white: carbon. Hydrogen atoms are not shown. Lead-centered octahedra are shaded gray. Structural parameters are given by Ref.~\cite{stoumposHighMembers2D2017b}. b)~Schematic view of the potential barriers created by the layers of butylammonium (BA) separating blocks of 5 layers of methylammonium lead iodide (MAPbI). c)~Attenuance spectra of the material in the vicinity of the band edge at temperatures between 315\,K and 11\,K during cooling, vertically offset for visual separation.}
    \label{fig:Overview}
\end{figure}

Two-dimensional Ruddlesden--Popper (RP) hybrid perovskites are characterized by the insertion of additional organic spacers between tunable-thickness layers of perovskite material. They are described by the general form A$^\text{s}_2$A$_{n-1}$M$_n$X$_{3n+1}$ with spacer $\rm{A^s}$ and perovskite AMX$_{3}$ (M: metal, X: halide), where the parameter $n$ indicates the number of metal polyhedra between spacer layers. Figure~\ref{fig:Overview}(a) shows a view of the crystal structure of (BA)$_2$(MA)$_{n-1}$Pb$_n$I$_{3n+1}$ (BA: butylammonium $\mathrm{CH_3(CH_2)_3NH_3}$, MA: methylammonium $\mathrm{CH_3NH_3}$, $n = 5$), with the lead octahedra and butylammonium spacers visible~\cite{ruddlesdenNewCompoundsK2NIF41957, kaganOrganicInorganicHybridMaterials1999, blanconExtremelyEfficientInternal2017, stoumposHighMembers2D2017b}. For this study, large single crystals of $\rm{(BA)_2(MA)_{4}Pb_4I_{16}}$ (2-3\,mm) were prepared on glass substrates by the previously reported kinetic-controlled space confinement method~\cite{houSynthesis2DPerovskite2024}. Precise control of reaction kinetics allows for the slow intercalation of precursor ions, gradually increasing the $n$-order of the RPP until the desired value is reached. 

The insulating barrier introduced by the butylammonium spacers is illustrated in Fig.\,\ref{fig:Overview}(b), and imposes sufficient confinement in the stacking direction to increase the 16\,meV exciton binding energy of the 3D perovskite $\rm{MAPbI_3}$ by an order of magnitude for $n$ = 2, 3, 4, and 5 and nearly 30 times for $n=1$ at liquid helium temperatures~\cite{blanconScalingLawExcitons2018, miyataDirectMeasurementExciton2015, yangUnravelingExcitonBinding2017}. Zero-field attenuance spectra at multiple temperatures in Fig.\,\ref{fig:Overview}(c) illustrate the broadening of spectroscopic features at elevated temperatures, as well as the disappearance of the feature marking the interband transition ($\sim$2\,eV) at temperatures above 11\,K, which greatly complicate the measurements of exciton binding energy by purely optical methods~\cite{blanconExtremelyEfficientInternal2017, hansenMeasuringExcitonBinding2024a}.

\begin{figure*} % Originally a one-column figure, the diagram in a) was way too small so now it spans
    \centering
    \includegraphics[width=1\linewidth]{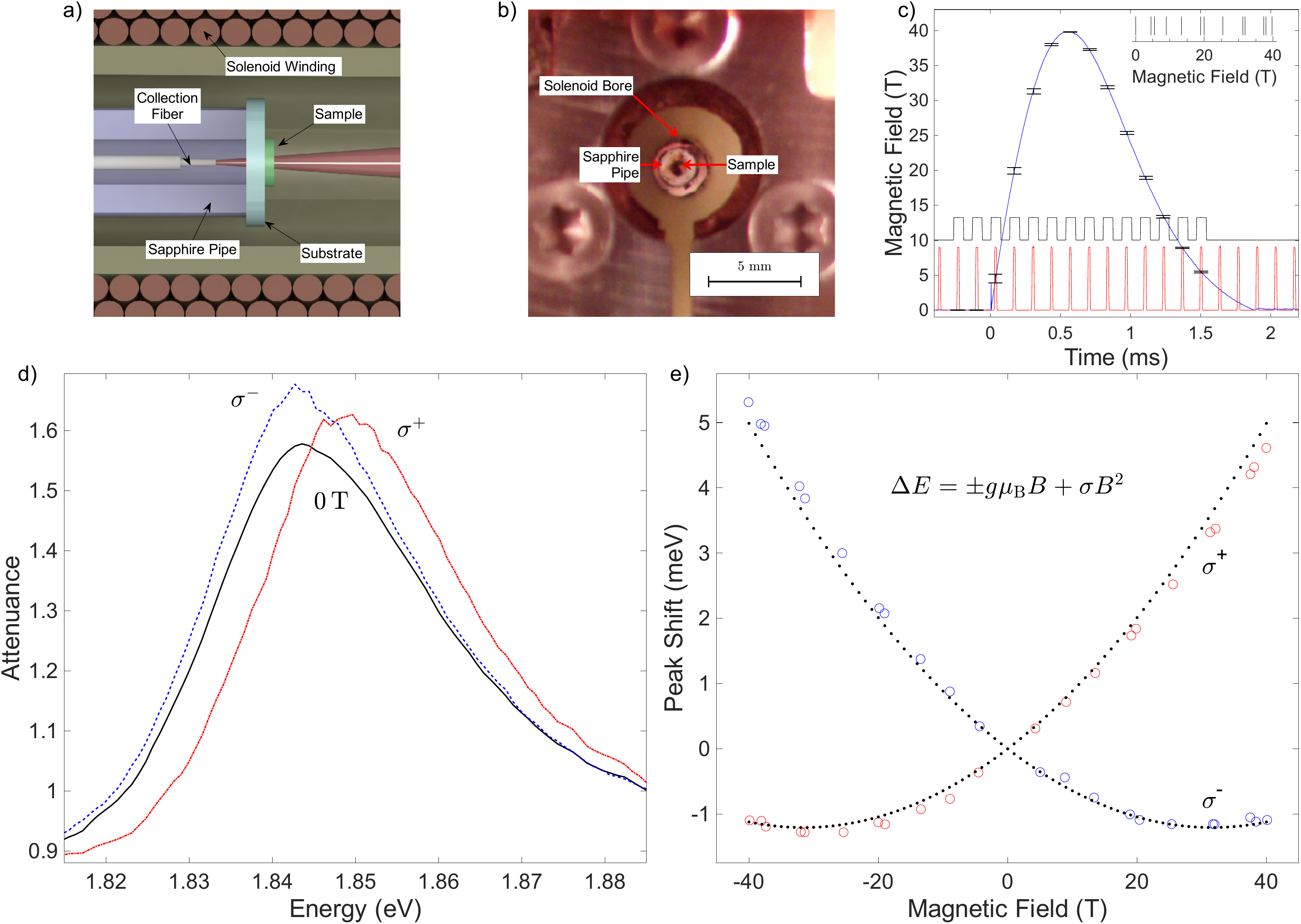}
    \caption{Experimental overview for RAMBO-II (40\,T). a)~Model view of the magnet interior. Visible are the coil windings, sapphire pipe, optical fiber, and mounted sample. b)~Camera view along the optical axis down the magnet bore, in which the sample is visible at the end of the sapphire pipe. c)~Timing chart for synchronization of measurement equipment. From top to bottom: photodiode, sync pulse from delay generator, trigger signal from hand switch to the capacitor bank controller, `Trigger Out' gate signal starting with the hand trigger and lasting until the end of discharge, magnetic field trace from Rogowski coil beginning after a calibrated delay $\Delta t_\text{D}$ to leave room for a zero-field exposure, frame-by-frame trigger signals for frame-transfer EMCCD detector operating in kinetics mode. d)~Attenuance spectra of the exciton peak at zero field and 40\,T in right- and left-circular polarization. Data were taken in pairs of zero field and high field during one pulse event and paired together. e)~Exciton peak shift extracted from paired zero and high field data fitted to Equation~(\ref{eq:shift}) with $g = 2.6$ and $\sigma = 1.2\,\mathrm{\upmu eV/T^2}$ for this dataset.}
    \label{fig:RAMBO}
\end{figure*}

\subsection{\label{sec:pulsedMagnet}The Rice Advanced Magnet with Broadband Optics}%The Rice Advanced Magnet with Broadband Optics}
The measurements of diamagnetic shift as a function of temperature were carried out with two generations of the Rice Advanced Magnet with Broadband Optics: RAMBO-I (nominally 30\,T) and RAMBO-II (nominally 50\,T). The temperature-dependent magneto-attenuance measurements up to 25\,T were conducted using RAMBO-I~\cite{noeTabletopRepetitivePulsed2013, noeSingleshotTerahertzTimedomain2016, baydinTimedomainTerahertzSpectroscopy2021, tayMagnetoopticalSpectroscopyRAMBO2022}. The near-infrared/visible free-space optical experiment employed a supercontinuum laser and a diffraction-grating spectrograph. Detailed cryostat configuration, sample mounting, and magnet operation can be found in previous reports, with the magnet timing protocol similar to Ref.~\cite{noeTabletopRepetitivePulsed2013} and differing only marginally due to faster measurement systems. The repetition rate of the NKT Photonics SuperK EXTREME EXU-6 supercontinuum laser at 78\,MHz greatly exceeded the framerate of the CCD and the sampling speed of analog components involved, so the light source was effectively a continuous wave in time. An optical chopper produced light pulses at 5\% duty cycle. Multiple analog signals, namely the generated magnetic field and the chopped light pulse, were recorded during the experiment to verify experiment synchronization. The latter was measured by a photodiode while the former was measured twice, first by a current transformer, which measures pulse current at the capacitor bank, and second by a pickup coil inside the solenoid. Signal from the photodiode also triggered a delay generator to provide clean digital edges for synchronization. The probe beam was circularly polarized using a linear polarizer and a quarter-wave plate. During each magnetic field pulse, two frames of data were taken during adjacent chopper pulses, and the capacitor bank was timed to discharge after a specified delay so that a zero-field spectrum can be acquired immediately before each high-field data point taken during the pulse to minimize any laser drift effects. Data was taken for both light polarizations and both field polarities.
%Figure \ref{fig:RAMBO}(c) summarizes this scheme.

Additional data of an impure sample up to 40\,T was acquired with the RAMBO-II IR-VIS setup, the 50\,T-capable expansion to the RAMBO project. To reach such extreme fields reliably, this new setup has a solenoid with a much smaller free bore of 3\,mm comprising 13 layers of 11 turns each. Reducing the bore volume reduces the total magnetic flux and thus the total energy for a given peak flux density. Hence, we chose to collect the optical signal with a fiber placed near the sample instead of by a completely free-space optical system; see Fig.~\ref{fig:RAMBO}(a-b). Additionally, we have to remove the pickup coil due to the reduced clearance between the sample rod and the high-voltage elements after the field is calibrated. 

Even with the reduction in bore and a reduced operating capacitance (2\,mF compared to 5.6\,mF), the solenoid generated a substantial amount of heat during each discharge, which must be dissipated. Each 30\,T shot required 20 minutes of cooldown compared to the RAMBO-I design with 5-minute cooldown (due to immersion of the coil in liquid nitrogen). Thus, there is a significant incentive to improve the efficiency of data collection. To that end, the electron-multiplied charge-coupled device (EMCCD) was operated in kinetics mode to increase the amount of data acquired for each discharge. A custom low-duty-cycle optical chopper blade was designed to increase the blanking interval between light pulses and protect the CCD from smearing during the frame transfer process. The resulting scheme more closely resembles that previously described in Ref.~\cite{baydinTimedomainTerahertzSpectroscopy2021}. 

Uncertainty in the sampled magnetic field was derived from the distribution of sampled fields weighted by pulse intensity during the sampling window, summarized in Fig.\,\ref{fig:RAMBO}(c). A total of 14 frames are recorded, but the first is smeared from imperfect pretrigger cleaning, and the fourteenth is overexposed by being unmasked during analog-to-digital conversion, so only one zero-field and eleven high-field frames are usable. In this manner, measurements can be conducted at a range of fields up to the peak field at satisfactory density for a complete experiment (pulses and field sampling shown in Fig.\,\ref{fig:RAMBO}(c)). Data taken in all four permutations of probe light circular polarization and field polarity (Fig.~\ref{fig:RAMBO}(e)) was then fitted simultaneously using the equation
\begin{align}
\Delta E(B) = \pm g\mu_\text{B} B + \sigma B^2,
\label{eq:shift}
\end{align}
where $B$ is the applied magnetic field, $\Delta E(B)$ is the $B$-induced shift in energy, $g$ is the $g$-factor, $\mu_\text{B}$ is the Bohr magneton, and $\sigma$ is the diamagnetic shift coefficient.  The first term is the Zeeman splitting, and the second term is the diamagnetic shift. Additional details on the operation of the high-magnetic field experiment are provided in Appendix~\ref{sec:Synch}. 

%Zero-field and high-field measurements were taken for each magnetic field pulse, and these spectra are compared to determine exciton peak shift at high field relative to the most proximal zero field measurement (see Fig.~\ref{fig:RAMBO}(d) for characteristic spectra). 

%% file: Sections/3_results.tex
\begin{figure*}
    \centering
    \includegraphics[width=1\linewidth]{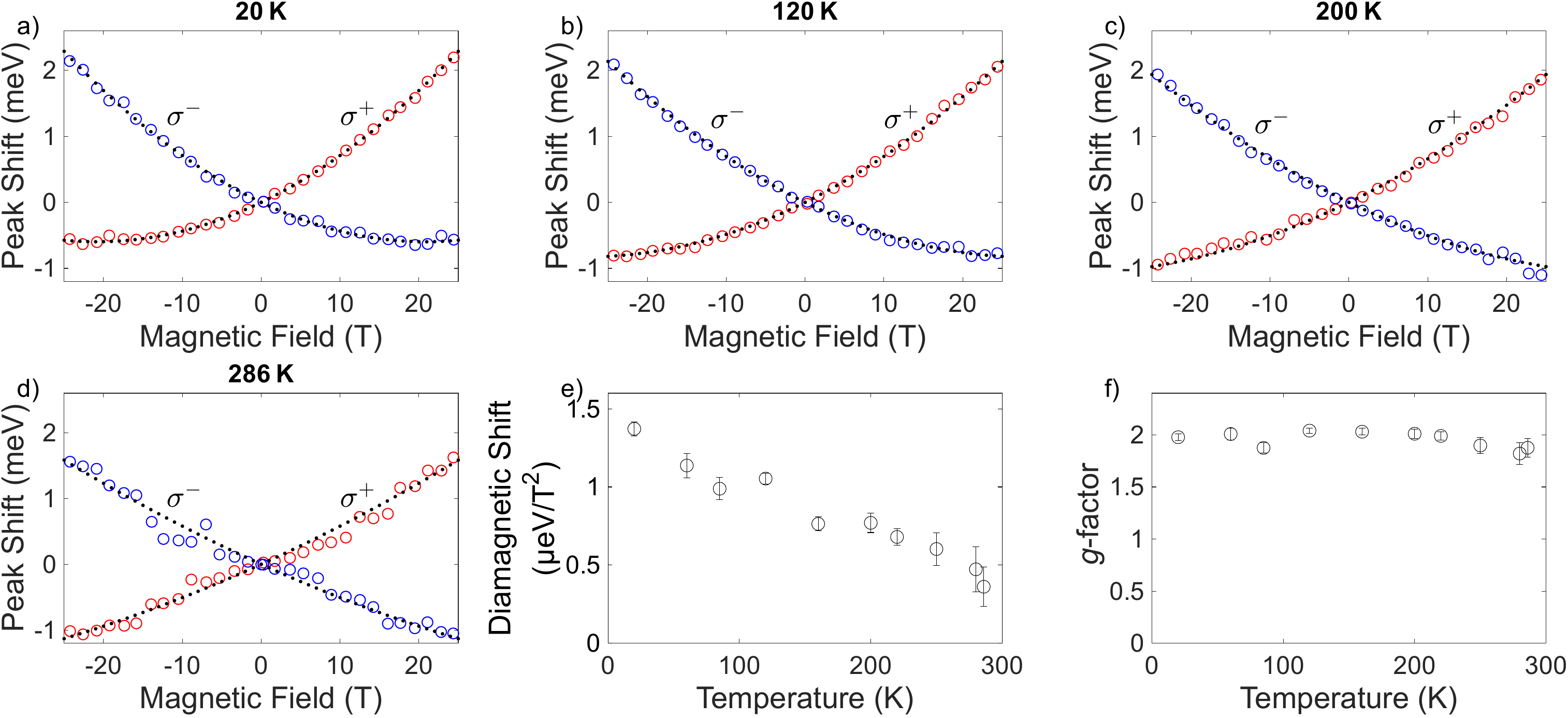}
    \caption{Field-dependent exciton peak shift at a)~20\,K, b)~120\,K, c)~200\,K, and d)~286\,K measured and fitted in the same manner as Fig.~\ref{fig:RAMBO}(e). e)~Diamagnetic shift coefficient obtained at 10 temperatures is inversely correlated with temperature. Data and fitting for the additional temperatures are provided in Appendix Figure~\ref{fig:DetailedResults}. f)~The $g$-factor of the excitons shows no systematic temperature dependence.
} 
    \label{fig:Results}
\end{figure*}

In addition to the 40\,T data presented in Fig.~\ref{fig:RAMBO}, we collected the attenuance spectra and extracted the peak shift of the $1s$ exciton state at positive and negative magnetic fields, and in right- and left-circular polarization collected at ten temperatures between 20\,K and 286\,K, as described in Section~\ref{sec:pulsedMagnet}. Figure~\ref{fig:Results}(a-d) shows exciton peak shift as a function of magnetic field for representative temperatures of 20\,K, 120\,K, 200\,K, and 286\,K. Data from all temperatures are presented in further detail in Appendix Figure~\ref{fig:DetailedResults}.

At low temperatures (20\,K), the diamagnetic shift coefficient, $\sigma$, reaches a maximum observed value of 1.37~$\pm$~0.05\,$\upmu$eV/T$^2$, recognizable in Fig.\,\ref{fig:Results}(a) by the quadratic character of the peak shift in each of the right- and left-handed traces. 
The average value of the peak shift probed by right- and left- circularly polarized light moves notably up at larger fields, contrasting with the small quadratic character at higher temperatures as in Fig.\,\ref{fig:Results}(d). Reduced values of diamagnetic shift coefficient are measured with increasing temperatures, down to a minimum of 0.36~$\pm$~0.13\,$\upmu$eV/T$^2$ at 286\,K, as shown in Figure\,\ref{fig:Results}(e). The value of the $g$-factor was observed in the range 1.82~$\pm$~0.10 to 2.04~$\pm$~0.03 with no temperature dependence as reflected by a linear fit to the data in Fig.\,\ref{fig:Results}(f) with constant 2.0~$\pm$~0.1, slope $-$0.4~$\pm$~0.6~$\times$~10$^{-3}$\,K$^{-1}$, and $R^2 = 0.24$. 

%Data presented in Figure \ref{fig:RAMBO}(d-e) was taken at 130\,K in RAMBO-II under liquid nitrogen transfer. The sample used for this measurement was polyphasic, with the characteristic peaks of the $n = 4, 5, 6,$ and 7 phases observable at low temperature while only $n = 5$ was strong enough to observe at room temperature.
%(additional spectra in Supplemental Material Figure \textcolor{red}{XXXX}).

%This polyphasic sample showed magnetoexcitonic peak shift of the $n = 5$ phase with $g = 2.6 \pm 0.2$ and $\sigma = 1.2 \pm 0.2\, \mathrm{\mu eV/T^2}$, both values having increased with respect to the phase-pure sample at comparable temperature ($g = 2.04 \pm 0.03$ and $\sigma = 1.05 \pm 0.03\, \mathrm{\mu eV/T^2}$ at 120\,K).

%% file: Sections/4_discussion.tex
Such strong temperature-dependence of diamagnetic shift is unexpected for most semiconductor materials. In ideal 2D and 3D hydrogen models, the diamagnetic shift is related to material parameters that usually do not change notably as a function of temperature: $\sigma_\text{3D} = 4\pi^2\epsilon^2\hbar^4/e^2(\mu^*)^3$ ~\cite{dyksikUsingDiamagneticCoefficients2022, congExcitonsMagneticFields2018}, where $\epsilon$ is the dielectric constant, $\mu^*$ is the reduced effective mass of the electron--hole pair, and $\hbar$ is the reduced Planck constant. The equation can also be rewritten as a function of exciton radius, $a^*_\text{B,3D}$, as $\sigma_\text{3D} = e^2(a^*_\text{B,3D})^2/4\mu^*$, or a function of exciton binding energy, $E_\text{B,3D}$, as $\sigma_\text{3D} = \hbar^2e^2/8(\mu^*)^2E_\text{B,3D}$. Although existing literature has demonstrated that the excitons in the $n$ = 1--5 compositions of (BA)$_2$(MA)$_{n-1}$Pb$_n$I$_{3n+1}$ deviate from both the 2D and 3D hydrogen atom model~\cite{blanconScalingLawExcitons2018}, one may still expect that the mixed dimensionality of hybrid perovskites only modifies the prefactors of these equations~\cite{miuraPhysicsSemiconductorsHigh2007, congExcitonsMagneticFields2018}. Here we kept $\mu^*$ constant because the optical band gap of our material appears to be stable within a few percent, even during a phase transition below room temperature; see Fig.\,\ref{fig:PeakTempShift}. Other experimental evidence provides a justification for this assumption in hybrid perovskites, for which $\mu^*=0.104m_0$, where $m_0 = 9.11 \times 10^{-31}$\,kg, has been measured via high-field magneto-optical spectroscopy at 2\,K and 160\,K in both the orthorhombic and tetragonal structural phases of MAPbI~\cite{miyataDirectMeasurementExciton2015,yangUnravelingExcitonBinding2017}. Under these assumptions, the exciton binding energy can be expressed as a function of diamagnetic shift: $E_\text{B,3D} = \hbar^2e^2/[8(\mu^*)^2\sigma_\text{3D}]$. This inverse proportionality relationship between the binding energy and diamagnetic shift coefficient is the same in the 2D and 3D hydrogen atom models, where the expressions differ only by a constant factor. The change in measured diamagnetic shift coefficient from 1.37~$\pm$~0.05\,$\upmu$eV/T$^2$ at 20\,K to 0.36~$\pm$~0.13\,$\upmu$eV/T$^2$ at 286\,K, shown in Fig.\,\ref{fig:Results}(e), corresponds to an increase in binding energy of 3.8 times, suggesting a room-temperature exciton binding energy of nearly half an electron volt, much larger than the previously measured value 125~$\pm$~29\,meV at 4\,K~\cite{blanconScalingLawExcitons2018}.

The other possible temperature-dependent variable of the material is the dielectric constant, since polar organic molecules are known to have a strong temperature-dependent dielectric response from rotation and libration modes~\cite{solsonaDielectricPropertiesTen1982}. Qualitatively, thermal fluctuations reduce the ability of polar molecules to screen external fields by alignment at higher temperatures. The reduced dielectric function may in principle increase the exciton binding energy. It is rather difficult to quantify this effect within the scope of this work because the dielectric function is also strongly frequency- and momentum-dependent, and selecting a proper value in excitonic calculations to account various screening contributions is nontrivial~\cite{knoxTheoryExcitons1963, umariInfraredDielectricScreening2018, fenebergManyelectronEffectsDielectric2016, hansenMeasuringExcitonBinding2024a, songDeterminationDielectricFunctions2021}. Moreover, the ratio between exciton radius and the thickness of the unit cell may also affect the applicability of the 2D and 3D hydrogen atom models. Overall, we expect that more systematic studies of the temperature-dependent dielectric function in the infrared frequencies, nonlinear spectroscopy measurements of optically forbidden exciton levels, and advanced computational methods may help solve the puzzle.

%% file: Sections/5_conclusion.tex
In summary, we present temperature-dependent magnetoattenuance spectroscopy of the $n = 5$ member of the 2D Ruddlesden-Popper hybrid perovskite (BA)$_2$(MA)$_{n-1}$Pb$_n$X$_{3n+1}$ at magnetic fields up to 40~Tesla. We found a significant reduction in diamagnetic shift coefficient from 1.4\,$\upmu$eV/T$^2$ at 20\,K to 0.4\,$\upmu$eV/T$^2$ near room temperature with no accompanying change in $g$-factor. For an excitonic system following the hydrogen atom model, this would suggest that the room-temperature exciton binding energy exceeds the already large low-temperature value of 125\,meV \cite{blanconScalingLawExcitons2018} by a factor of 3.8 to reach 475\,meV. Extrapolation from the diamagnetic shift coefficient to the binding energy is not direct, but future experiments of the dielectric function and more direct measurements of the binding energy would provide clarification. The presence of high-binding-energy excitons would require the exploitation of edge states~\cite{blanconExtremelyEfficientInternal2017} or the engineering of heterostructures \cite{sidhikDeterministicFabrication3D2022, hanSurface2DBulk3DHeterophased2020} to promote exciton dissociation into free carriers in photovoltaic devices. The fact that 2D RP-perovskites have already been integrated into high-efficiency devices in spite of such large apparent room-temperature exciton binding energies speaks to the efficiency of these dissociation mechanisms. Further investigations are needed to clarify the origin of the unconventional temperature dependence of the exciton binding energy in this material. %A potential opportunity to further enhance the photovoltaic performance of such materials would be to reduce the excitonic binding energies or apply these mechanisms to systems with lower binding energies.

%% file: Sections/A_Experiment.tex
\begin{figure}
    \centering
    \includegraphics[width=1\linewidth]{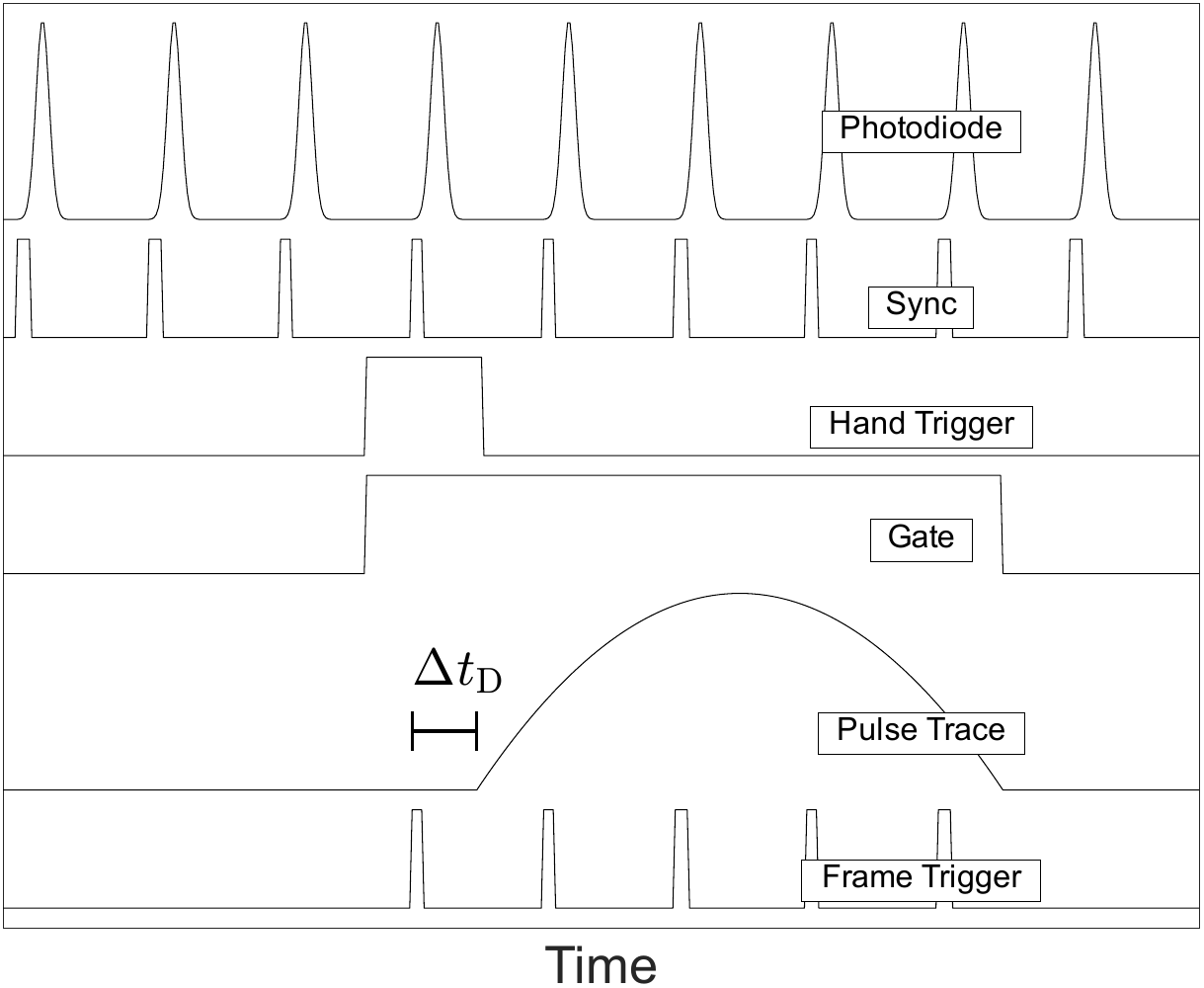}
    \caption{Signals involved in the synchronization of a pulsed magnet experiment.}
    \label{fig:Timing}
\end{figure}

Figure \ref{fig:Timing} shows a summary of the various signals involved in synchronizing a high-field pulsed magnet experiment. Timing is built around the chopped light pulses as read by a photodiode. Each light pulse triggers a delay generator to produce a digital TTL pulse slightly before the next chopper period, providing a consistent digital edge for all equipment to reference. When the discharge signal is sent to the capacitors from the user interface or hand trigger, the controller waits to initiate the discharge sequence until the next reference edge provided by the delay generator. From that point until the end of the discharge, the controller sets a new gate signal, and the CCD trigger signal is the logical AND of this gate and the digital pulses from the delay generator. Use of this gate prevents the spectroscopy equipment from being triggered outside of the discharge event. A delay $\Delta t_\text{D}$ allows for a zero-field transmission measurement to be made immediately before the high-field measurements.

%% file: Sections/A_ZeroField.tex
\begin{figure}
    \centering
    \includegraphics[width=1\linewidth]{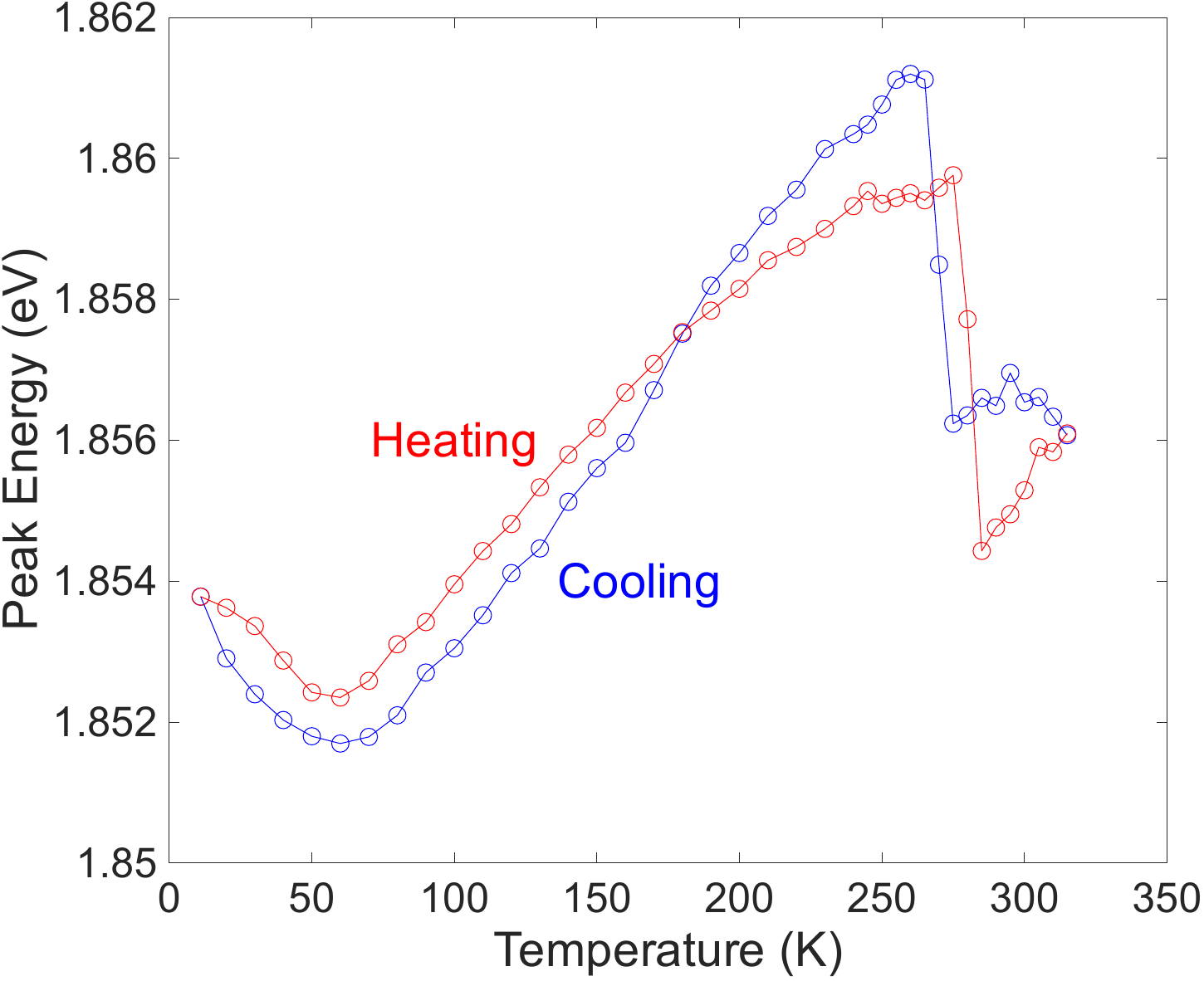}
    \caption{Temperature-dependent position of the exciton peak in $\rm{(BA)_2(MA)_{4}Pb_4I_{16}}$, warming and cooling.}
    \label{fig:PeakTempShift}
\end{figure}

The temperature-dependent position of the exciton peak in the same sample used for 25\,T temperature-dependent magnetospectroscopy was measured at finer temperature intervals while heating and cooling from 11 to 315\,K. A steep jump is observed near 280\,K (Figure \ref{fig:PeakTempShift}). 

% \section{Polyphasic Sample}

% \renewcommand{\thefigure}{\thesection.\arabic{figure}}
% \setcounter{figure}{0}

% \begin{figure}[h]
%     \centering
%     \includegraphics[width=1\linewidth]{Polyphasic.pdf}
%     \caption{Zero-field attenuance spectrum of a polyphasic $\rm{BA_2MA_{n-1}Pb_nI_{3n+1}}$ sample at 130\,K. Most prominent peak belongs to the $n=5$ phase.}
%     \label{fig:Polyphasic}
% \end{figure}

%% file: Sections/A_DetailedResults.tex
\begin{figure*}
    \centering
    \includegraphics[width=1\linewidth]{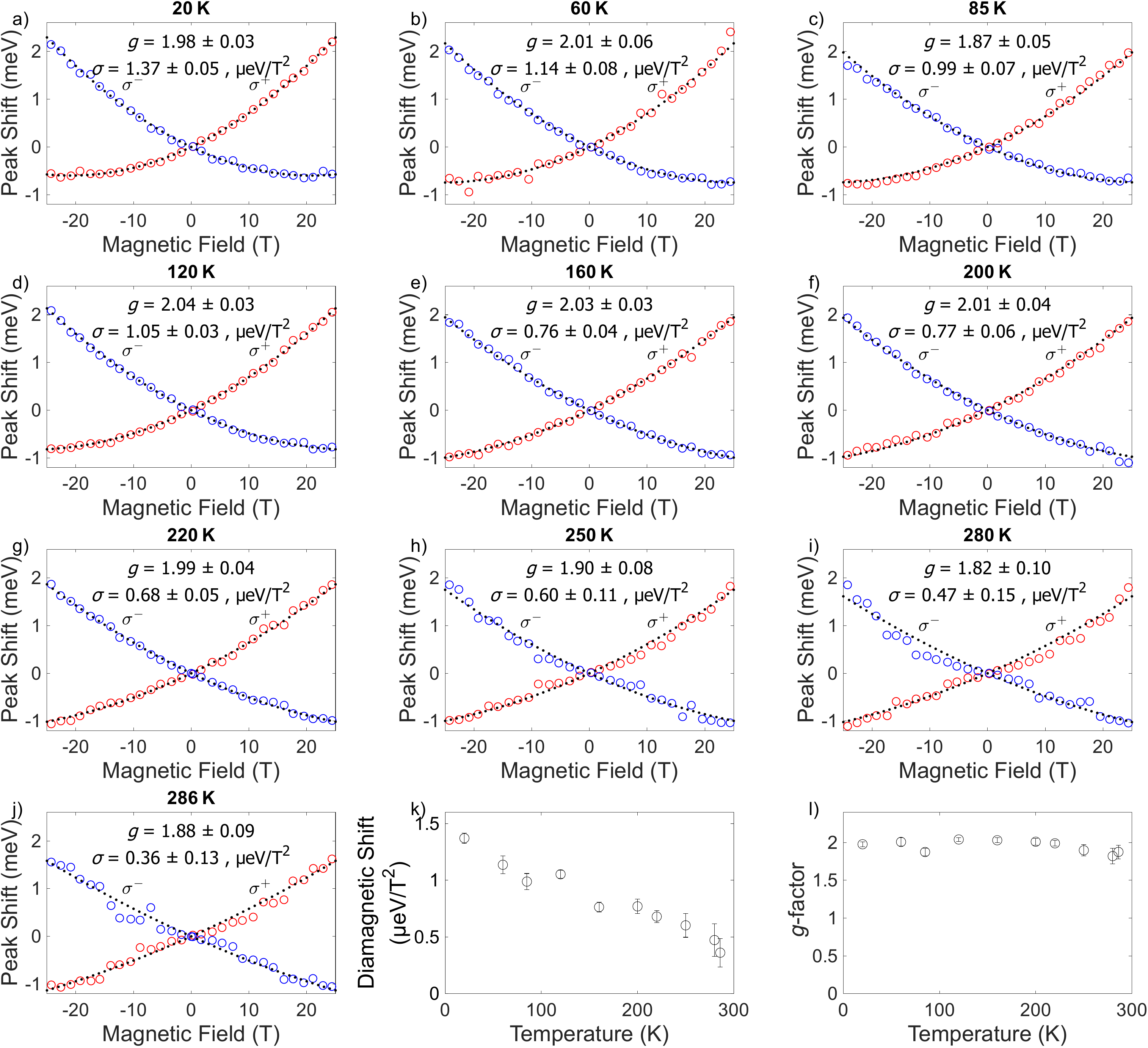}
    \caption{Enumerated and specified field-dependent exciton peak shift for all temperatures (a-j) at which data were collected. k)~Diamagnetic shift, reproduced from \ref{fig:Results}(e). l)~$g$-factor for these same data, reproduced from \ref{fig:Results}(f).}
    \label{fig:DetailedResults}
\end{figure*}

Statistics for fits to experimental data presented for a selection of temperatures in Fig.\,\ref{fig:Results} are presented at every temperature in Fig.\,\ref{fig:DetailedResults}.

